\documentclass[twocolumn,showpacs,preprintnumbers,amsmath,amssymb]{revtex4}
\usepackage{graphicx}% Include figure files
\usepackage{dcolumn}% Align table columns on decimal point
\usepackage{bm}% bold math

\begin{document}

%\preprint{APS/123-QED}
\title{Quantum Pattern Retrieval by Qubit Networks with Hebb Interactions}% Force line breaks with \\

\author{M. Cristina Diamantini}
%\altaffiliation[Also at ]{Theory Division, CERN, CH-1211 Geneva 23, Switzerland}%Lines break automatically or can be forced with \\
%\author{Pasquale Sodano}%
\email{cristina.diamantini@pg.infn.it}
\affiliation{%
INFN and Dipartimento di Fisica, University of Perugia, via A. Pascoli, I-06100 Perugia, Italy
}%

\author{Carlo A. Trugenberger}
\email{ca.trugenberger@bluewin.ch}
\affiliation{%
Theory Division, CERN, CH-1211 Geneva 23, Switzerland\\
}%

%\affiliation{

\date{\today}% It is always \today, today,

\begin{abstract}
Qubit networks with long-range interactions inspired 
by the Hebb rule can be used as quantum associative memories. 
Starting from a uniform superposition, the unitary evolution generated by these interactions
drives the network through a quantum phase transition at a critical
computation time, after which ferromagnetic order guarantees that a measurement retrieves 
the stored pattern. The maximum memory capacity of these qubit networks is reached at a
memory density $\alpha =p/n =1$. 
\end{abstract}

\maketitle

%\section{Introduction}

Historically, the interest in neural networks \cite{neuralnetworks} has been driven by the desire to build machines 
capable of performing higher cognitive tasks, for which the sequential circuit paradigm of Babbage and von Neumann is not well
suited, like pattern recognition and categorization. 

On the other side, the last decade has seen the emergence of quantum mechanics as a powerful new paradigm
for computation \cite{review}. The focus of these developments has been mostly on the quantum circuit model,
in which a set of elementary universal quantum gates are sequentially applied on qubit registers, the root of the
interest lying mainly in the speed-up with respect 
to classical computation. 

In this paper we shall propose a different quantum information processing system, which is capable of
higher cognitive tasks as described above.
Instead of one- and two-qubit interactions that are switched on and off sequentially, we will consider 
fixed, long-range interactions inspired by neural networks, focusing
for concreteness on one particular model, the associative memory model of Hopfield \cite{hopfield}.  
The result is a fully-connected interacting network of qubits in which 
memories are encoded in the long-range interaction matrix: the corresponding Hamiltonian
generates a unitary evolution capable of memory retrieval and pattern recognition.  
Note that some of the most promising technologies for the
implementation of quantum information processing, like optical lattices \cite{optical} and arrays of quantum dots \cite{qudots}
rely exactly on similar collective phenomena. 

In mathematical terms, the simplest neural network model is a directed graph with the following properties:
\begin{enumerate}
\item{A state variable $n_i$ is associated with each node (neuron) $i$.}
\item{A real-valued weight $w_{ij}$ is associated with each link (synapse) $(ij)$ between two nodes $i$ and $j$.}
\item{A state-space-valued transfer function $f(h_i)$ of the synaptic potential $h_i = \sum_j w_{ij} n_j$ determines the dynamics of the network.}
\end{enumerate}
Two types of dynamical evolution have been considered: sequential or parallel synchronous. In the first case the
neurons are updated one at a time according to
\begin{equation}
n_i ( t + 1 ) = f \left( \sum_k w_{ik} n_k (t) \right) \ ,
\label{aa}
\end{equation}
while in the second case all neurons are updated at the same time. 
The simplest model is obtained when
neurons become binary variables taking only the values $n_i=\pm 1$ for all $i$ and the transfer function
becomes the sign function. This is the original McCullogh-Pitts \cite{pitts} neural network model, in 
which the two states represent quiescent and firing neurons.

The Hopfield model \cite{hopfield} is a fully-connected McCullogh-Pitts network 
in which the synaptic weights are symmetric quantities chosen according to the Hebb rule \cite{neuralnetworks}
\begin{equation}
w_{ij} = w_{ji} = {1\over {n-1}} \ \sum_{\mu =1}^p \xi_i^{\mu} \xi_j^{\mu} \ ,\qquad w_{ii}=0 \ .
\label{ab}
\end{equation}
Here $n$ is the total number of neurons and ${\bf \xi}^{\mu}$ are $p$ binary
patterns to be memorized ($\xi_i^{\mu} = \pm 1$) 

For $p/n < 0.138$ the network works as an associative memory, permitting recall of information
only on the basis of its content, without any information on storage location: 
the stored patterns are attractors for the dynamics (\ref{aa}) defined by the transfer function. For higher loading
factors $p/n$, however a phase transition to a spin glass phase \cite{spinglass}
impairs the ability to correctly recall memories.

The "quantization" of Hopfield networks can be carried out straightforwardly by substituting each neuron with a qubit i.e. a quantum degree
of freedom with a two-dimensional Hilbert space whose basis states can be labeled as $|0>$ and $|1>$. 
The only tricky point regards the treatment of interactions bewteen the qubits. In the classical model 
the dynamics (\ref{aa}) induced by the transfer function is fully deterministic and irreversible, which is not compatible with quantum mechanics. A first generalization that has been considered is that of stochastic neurons, in which the transfer function determines only the probabilities that the classical
state variables will take one of the two values: $n_i ( t + 1 ) = \pm 1$ with probabilities $f(\pm h_i(t))$, where $f$ 
must satisfy $f(h \to -\infty) = 0$, $f(h \to +\infty) = 1$ and $f(h)+f(-h)=1$. 

While this modification makes the dynamics probabilistic by introducing thermal noise, the
evolution of the network is still irreversible since the 
actual {\it values} of the neurons are prescribed after an update step. In quantum mechanics the evolution must be reversible and only the magnitudes of the {\it changes} in the neuron variables can be postulated. Actually, the dynamics must generate a {\it unitary}
evolution of the network. 

To this end we propose the following "transverse" Hamiltonian:
\begin{equation}
{\cal H} = J \sum_{ij}\ w_{ij} \sigma^y_i \sigma^z_j \ ,
\label{ac}
\end{equation}
where $\sigma ^k$, $k=x,y,z$ denote the Pauli matrices and $J$ is a coupling constant with the dimensions of mass (we use units $c=1, \hbar =1$).   
This generates a unitary evolution of the network:
\begin{equation}
|\psi (t)> = {\rm exp} (i{\cal H}t) \ |\psi_0> \ ,
\label{ad}
\end{equation}
where $|\psi_0> = |\psi (t=0)>$. Specifically,
we will choose as initial configuration of the network the uniform 
superposition of all computational basis states \cite{review}
\begin{equation}
|\psi _0> = {1\over \sqrt{2^n}}\ \sum_{x=0}^{2^n-1} |x> \ .
\label{ae}
\end{equation} 
This corresponds to a "blank memory" in the sense that all possible states have the same probability of being recovered upon measurement. 
In the language of spin systems this is a state in which all spins are aligned in the $x$ direction.  

Inputs $\xi^{\rm ext}$ can be accomodated by adding an external transverse magnetic field along the $y$ axis, 
i.e. modifying the Hamiltonian to 
\begin{equation}
{\cal H} = J \sum_{ij} w_{ij} \sigma^y_i \sigma^z_j + g \sum_{i} h_i^{\rm ext} \sigma^y_i \ , 
\label{af}
\end{equation}
where $h_i^{\rm ext} = \sum_{j} w_{ij} \xi_j^{\rm ext}$. 
This external magnetic field
can be thought of as arising from the interaction of the network with an additional "sensory" qubit register
prepared in the state $\xi^{\rm ext}$, the synaptic weights between the two layers 
being identical to those of the network self-couplings.  

Let us now specialize to the simplest case of one assigned memory ${\bf \xi}$ in which $w_{ij} = \xi_i \xi_j$. In the classical Hopfield
model there are two nominal stable states that represent attractors for the dynamics, the pattern ${\bf \xi}$ itself 
and its negative $-{\bf \xi}$. Correspondingly, the quantum dynamics defined by the Hamiltonian (\ref{ac})
and the initial state (\ref{ae}) have a $Z_2$ symmetry generated by $\prod_{i} \sigma^x_i$, corresponding to the 
inversion $|0> \leftrightarrow |1>$ of all qubits.  

Since the qubits of the network are fully connected and weakly interacting, the model can be expected to be exactly solvable
in the $n\to \infty$ limit by  a mean-field theory. Indeed, this is known to become exact for weak, long-range interactions
\cite{stat}. In the mean-field approximation operators are decomposed in a sum of
their mean values and fluctuations around it, $\sigma_i^k = <\sigma_i^k> + \left( \sigma_i^k - <\sigma_i^k> \right)$, and quadratic
terms in the fluctuations are neglected in the Hamiltonian. Apart from an irrelevant constant, this gives
\begin{eqnarray}
{\cal H}_{\rm mf} &&= J \sum_i \sigma^y_i \left( <h^z_i> + {g\over J} h_i^{\rm ext}\right) + \sigma^z_i <h^y_i> \ ,
\nonumber \\
< h^k_i > &&= \sum_j w_{ij} <\sigma^k_j> = \xi_i \ m^k \ ,
\label{aff}
\end{eqnarray}
where $m^k = (1/n)\ \sum_i <\sigma^k_i> \xi_i$ is the average overlap of the state of the network with the stored pattern.
This means that each qubit $i$ interacts with the average
magnetic field (synaptic potential) $< h^k_i >$ due to all other qubits: naturally, the correct values of these 
mean magnetic fields $<h^k_i>$ have to be determined self-consistently. 

To this end we compute the average pattern overlaps $m^k$ using the mean field Hamiltonian (\ref{aff}) to
generate the time evolution of the quantum state. This reduces to a sequence of factorized rotations
in the Hilbert spaces of each qubit, giving
\begin{eqnarray}
m^y &&= -{m^y \over \vert m\vert} \ {\rm sin} \ 2Jt \vert m\vert \ ,
\nonumber \\
m^z &&= {{m^z + (g/J)M^z} \over \vert m\vert} \ {\rm sin} \ 2Jt \vert m\vert \ ,
\label{afg}
\end{eqnarray}
where $\vert m\vert = \sqrt{\left( m^y \right)^2 + \left( m^z + (g/J)M^z \right)^2}$ and 
$M^z = (1/n) \ \sum_i \xi_i^{\rm ext} \xi_i$ is the average overlap of the external stimulus with the stored memory.

Before we present the detailed solution of these equations, 
let us illustrate the mechanism underlying the quantum associative memory. 
To this end we note that, for $g=0$, the pattern overlaps $m^y$ and $m^z$ in the two directions cannot
be simultaneously different from zero. As we show below, only $m^z \ne 0$ for $J>0$ (for $J<0$ the
roles of $m^y$ and $m^z$ are interchanged). In this case 
the evolution of the network becomes a sequence of $n$ rotations
\begin{equation}
\left( \begin{array}{cc}
{\rm cos}(Jt <h^z_i>) & {\rm sin}(Jt <h^z_i>) \\
-{\rm sin}(Jt <h^z_i>) & {\rm cos}(Jt <h^z_i>) \\
\end{array} \right)
\label{ag}
\end{equation}
in the two-dimensional Hilbert spaces of each qubit $i$. The rotation parameter is exactly the same synaptic potential
$h_i$ which governs the classical dynamics of the Hopfield model.  
When these rotations are applied on the initial state (\ref{ae}) they amount to a single update step
transforming the qubit spinors into 
\begin{equation}
{1\over \sqrt{2}} \ \left( \begin{array}{cc}
{\rm cos}(Jt <h^z_i>) + {\rm sin}(Jt <h^z_i>) \\
{\rm cos}(Jt <h^z_i>) - {\rm sin}(Jt <h^z_i>)\\
\end{array} \right) \ .
\label{ah}
\end{equation}
This is the generalization to quantum probability {\it amplitudes} of the probabilistic formulation of classical stochastic neurons.
Indeed, the probabilities for the qubit to be in its eigenstates $\pm 1$ after a time $t$, obtained by squaring the probability amplitudes, are given 
by $f(\pm <h^z>)$, where $f(<h^z>) = (1+{\rm sin}(2Jt <h^z>))/2$ has exactly 
the properties of an activation function (alternative to the Fermi function), at
least in the region $Jt < \pi/4$. In this correspondence, the effective coupling constant $Jt$ plays
the role of the inverse temperature, as usual in quantum mechanics. 

We shall now focus on a network without external inputs. In this case the equation for the average pattern overlaps has
only the solution $\vert m\vert=0$ for $0< Jt < 1/2$.  
For such small effective couplings (high effective temperatures), corresponding to weak synaptic connections or to
short evolution times, the network is unable to remember the
stored pattern. For $1/2 < Jt $, however, the solution $\vert m\vert =0$ becomes unstable, 
and two new stable solutions $m^z=\pm m_0$ appear. 
This means that the reaction of the
mean orientation of the qubit spinors against a small deviation $\delta m^z$ from the $\vert m\vert =0$ solution is larger than the deviation itself. 
Indeed, any so small external perturbation $(g/J)M^z$ present at the bifurcation time $t=1/2J$ is sufficient for the network evolution
to choose one of the two stable solutions, according to the sign of the external perturbation. The point $Jt =1/2$ 
represents a quantum phase transition \cite{sachdev} from an amnesia (paramagnetic) phase
to an ordered (ferromagnetic) phase in which the network has recall capabilities: the average pattern overlap $m^z$ is the corresponding order
parameter. In the ferromagnetic phase the original $Z_2$ symmetry of the model is spontaneously broken. 

%\begin{figure}
%\includegraphics[width=8cm]{qh1.eps}
%\caption{\label{fig:Fig1} The order parameter of quantum associative memories.}
%\end{figure}

For $Jt = \pi/4 $, the solution becomes $\vert m_0\vert =1$, which means that the network is capable of perfect recall
of the stored memory.
For $Jt > \pi/4$ the solution $m_0$ decreases slowly to 0 again. Due to the periodicity of the time evolution, however,
new stable solutions $m_0 = \pm 1$ appear at $Jt = (1+4n)\pi/4$ for every integer $n$. Also, for $Jt \ge 3\pi/4$, new solutions
with $m^y \ne 0$ and $m^z=0$ appear. These, however, correspond all to metastable states.
Thus, $t=\pi/4J$ is the ideal computation time for the network. 

The following picture of quantum associative memories emerges from the above construction. States of
the network are generic linear superpositions of computational basis states. The network is prepared in the state $|\psi _0>$ and
is then let to unitarily evolve for a time $t$. After this time the state of the network is measured, giving the result
of the computation. During the evolution each qubit updates its quantum state by a rotation that depends on the aggregated
synaptic potential determined by the state of all other qubits. These synaptic potentials are subject to large quantum fluctutations
which are symmetric around the mean value $<h^z>=0$. If the interaction is strong enough, any external disturbance will cause the 
fluctuations to collapse onto a collective rotation of all the network's qubits towards the nearest memory. 
 
We will now turn to the more interesting case of a finite density $\alpha =p/n$ of stored memories in the limit $n\to \infty$.
In this case the state of the network can have a finite overlap with several stored memories $\xi ^{\mu}$ simultaneously. As in the
classical case we shall focus on the most interesting case of a single "condensed pattern", in which the network uniquely recalls one
memory without admixtures.  
Without loss of generality we will chose this memory to be the first, $\mu =1$, omitting then the memory superscript on the
corresponding overlap $m$. Correspondingly we will consider external inputs so that only $M^{\mu=1} =M\ne 0$. 
For simplicity of presentation, we
will focus directly on solutions with a non-vanishing pattern overlap along the z-axis, omitting also the direction superscript $z$. 

In case of a finite density of stored patterns, one cannot neglect the noise effect due to the infinite number of memories. 
This changes (\ref{afg}) to
\begin{eqnarray}
m &&= {1\over n} \sum_i \ {\rm sin} \ 2Jt\left( m + {g\over J} M + \Delta_i\right) \ ,
\nonumber \\
\Delta_i &&= \sum_{\mu \ne 1} \xi_i^1 \xi_i^{\mu} m^{\mu} \ .
\label{am}
\end{eqnarray} 
As in the classical case we will assume that $\{ \xi_i^{\mu}\}$ and $\{ m^{\mu}, \mu \ne 1\}$ are all independent random variables
with mean zero and we will denote by square brackets the configurational average over the distributions of these random variables.
As a consequence of this assumption, the mean and variance of the noise term are given by $[\Delta_i] = 0$ and
$[\Delta_i^2] = \alpha r$, where 
\begin{equation}
r = {1\over \alpha} \ \sum_{\mu \ne 1} \left[ \left( m^{\mu} \right)^2 \right]
\label{an}
\end{equation}
is the spin-glass order parameter \cite{hopfield}. According to the central limit theorem one can now
replace $n^{-1} \sum_i$ in (\ref{am}) by an average over a Gaussian noise,
\begin{equation}
m = \int {dz\over \sqrt{2\pi}} \ {\rm e}^{-z^2 \over 2} \ {\rm sin} \ 2Jt \left( m + {g\over J} M + \sqrt{\alpha r} z \right) \ .
\label{ao}
\end{equation}

The second order parameter $r$ has to be evaluated self-consistently by a similar procedure starting from the equation
analogous to eq. (\ref{am}) for $\mu \ne 1$. In this case one can use $m^{\mu } \ll 1$ for $\mu \ne 1$ to expand the transcendental function
on the right-hand side in powers of this small parameter, which gives
\begin{eqnarray}
v &&= \int {dz\over \sqrt{2\pi}} \ {\rm e}^{-z^2 \over 2} 
\ {\rm sin}^2 \ 2Jt \left( m + {g\over J} M + \sqrt{\alpha r} z \right) \ ,
\nonumber \\
x &&= \int {dz\over \sqrt{2\pi}} \ {\rm e}^{-z^2 \over 2} \ {\rm cos} \ 2Jt \left( m + {g\over J} M + \sqrt{\alpha r} z \right) \ ,
\label{aq}
\end{eqnarray}
where $v = (1-2Jtx)^2 r$. 
Solving the integrals gives finally the following coupled equations for the two order parameters $m$ and $r$:
\begin{eqnarray}
m &&= {\rm sin} \ 2Jt \left( m+{g\over J} M \right) \ {\rm e}^{-2 (Jt)^2 \alpha r} \ ,
\nonumber \\
r &&= {1\over 2} \ {{1-{\rm cos} \ 4Jt \left( m+{g\over J} M\right) \ {\rm e}^{-8 (Jt)^2 \alpha r}} \over
{\left( 1-2Jt {\rm cos} \ 2Jt \left( m+{g\over J} M\right) \ {\rm e}^{-2 (Jt)^2 \alpha r} \right)^2}} \ .
\label{ar}
\end{eqnarray}

In terms of these order parameters one can distinguish three phases of the network. First of all the value of $m$ determines
the presence ($m>0$) or absence ($m=0$) of ferromagnetic order (F). If $m=0$ the network can be in a paramagnetic phase (P) if also
$r=0$ or a quantum spin glass phase (SG) if $r>0$. The phase structure resulting from a numerical solution of the coupled equations
(\ref{ar}) for $g=0$ is shown in Fig. 1. 

\begin{figure}
\includegraphics[width=8cm]{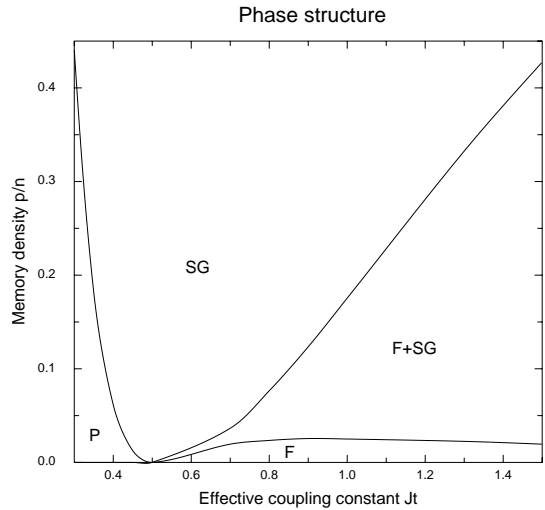}
\caption{\label{fig:Fig1} The phase structure of quantum associative memories with finite density of stored patterns. P, F and SG 
denote (quantum) paramagnetic, ferromagnetic and spin-glass phases, respectively. F + SG denotes a mixed phase in which the 
memory retrieval solution is only locally stable.}
\end{figure}

For $\alpha < 0.025$ the picture is not very different from the single memory case. For large enough computation times
there exists a ferromagnetic phase in which the $m=0$ solution is unstable and the network has recall capabilities. The only
difference is that the maximum value of the order parameter $m$ is smaller than 1 (recall is not perfect due to noise) 
and the ideal computation time $t$ at which the maximum is reached depends on $\alpha$. For $0.025 <\alpha <1.000$
instead, ferromagnetic order coexists as a metastable state with a quantum spin glass state. This means that ending up in the 
memory retrieval solution depends not only on the presence of an external stimulus but also on its magnitude; in other words, the
external pattern has to be close enough to the stored memory in order to be retrieved. 
For $1<\alpha $ all retrieval capabilities are lost and the network will be in a quantum spin glass state for all computation
times (after the transition from the quantum paramagnet). $\alpha =1$ is thus the maximum memory capacity of this quantum network.
Note that $\alpha = 1$ corresponds to the maximum possible number of linearly independent memories. 
For memory densities smaller but close to this maximum value, however,
the ferromagnetic solution exists only for a small range of effective couplings centered around
$Jt \simeq 9$: for these high values of $Jt$ the quality of pattern retrieval is poor, the value of the order parameter $m$ being
of the order 0.15-0.2. Much better retrieval qualities are obtained for 
smaller effective couplings: e.g. for $Jt=1$ the order parameter is larger than 0.9 (corresponding to an error rate smaller than 5\%) for
memory densities up to 0.1. In this case, however the maximum memory density is 0.175, comparable with the classical result of the Hopfield model. 

In summary, we have introduced a new model of quantum information processing that is capable of content association by quantum
self-organization. The basic idea is to encode information in Hebb-type qubit interactions and to let the corresponding quantum
evolution amplify the desired result, given an input, a principle that lends itself to many generalizations.

\end{document}